\documentclass[aps,prl,reprint,superscriptaddress]{revtex4-1}
\usepackage{graphicx}
\usepackage{mathrsfs}
\usepackage{gensymb}
\usepackage{xcolor}
\usepackage{ulem}
\usepackage[latin9]{inputenc}
\usepackage{textcomp}
\usepackage[bookmarks=true,bookmarksnumbered=false,bookmarksopen=false,
 breaklinks=false,pdfborder={0 0 0},pdfborderstyle={},backref=false,colorlinks=true]
 {hyperref}
\hypersetup{urlcolor=blue,linkcolor=black}

\usepackage{amsmath,amsfonts,amssymb}

\newcommand{\parallelsum}{\mathbin{\!/\mkern-5mu/\!}}

\providecolor{PIadded}{rgb}{1,0,0}
\providecolor{PIdeleted}{rgb}{0.6,0.6,0.8}
\begin{document}

\title{Realization of a Type-II Nodal-Line Semimetal in Mg$_3$Bi$_2$}

\author{Tay-Rong~Chang*}
\affiliation {Department of Physics, National Cheng Kung University, Tainan, 701, Taiwan}
\author{Ivo~Pletikosic*}
\affiliation {Department of Physics, Princeton University, Princeton, NJ 08544, USA}
\author{Tai~Kong*}
\affiliation {Department of Chemistry, Princeton University, Princeton, NJ 08544, USA}
\author{Guang~Bian}
\affiliation {Department of Physics and Astronomy, University of Missouri, Columbia, MO 65211, USA}
\author{Angus~Huang}
\affiliation {Department of Physics, National Tsing Hua University, Hsinchu 30013, Taiwan}
\author{Jonathan~Denlinger}
\affiliation {The Advanced Light Source, Lawrence Berkeley National Laboratory, Berkeley, CA 94720, USA}
\author{Satya~K.Kushwaha}
\affiliation {Department of Chemistry, Princeton University, Princeton, NJ 08544, USA}
\author{Boris~Sinkovic}
\affiliation {Department of Physics, University of Connecticut, Storrs, CT 06269, USA}
\author{Horng-Tay~Jeng}
\affiliation {Department of Physics, National Tsing Hua University, Hsinchu 30013, Taiwan}
\affiliation {Institute of Physics, Academia Sinica, Taipei 11529, Taiwan}
\author{Tonica~Valla}
\affiliation {Condensed Matter Physics and Materials Science, Brookhaven National Laboratory, Upton, NY 11973, USA}
\author{Weiwei~Xie}
\affiliation {Department of Chemistry, Louisiana State University, Baton Rouge, LA 70803, USA}
\author{Robert~J.~Cava}
\affiliation {Department of Chemistry, Princeton University, Princeton, NJ 08544, USA}


\date{\today}

\begin{abstract}

Nodal-line semimetals (NLSs) represent a new type of topological semimetallic beyond Weyl and Dirac semimetals in the sense that they host closed loops or open curves of band degeneracies in the Brillouin zone. Parallel to the classification of type-I and type-II Weyl semimetals, there are two types of NLSs. The conventional NLS phase, in which the two bands forming the nodal line have opposite signs for their slopes along any direction perpendicular to the nodal line, has been proposed and realized in many compounds, whereas the exotic type-II NLS is very rare. Our first-principles calculations show that Mg$_3$Bi$_2$ is a material candidate that hosts a single type-II nodal loop around $\Gamma$. The band crossing is close to the Fermi level and the two crossing bands have the same sign in their slopes along the radial direction of the loop, indicating the type-II nature of the nodal line. Spin-orbit coupling generates only a small energy gap ($\sim$35 meV) at the nodal points and does not negate the band dispersion of Mg$_3$Bi$_2$ that yields the type-II nodal line. Based on this prediction we have synthesized Mg$_3$Bi$_2$ single crystals and confirmed the presence of the type-II nodal lines in the material. Our angle-resolved photoemission spectroscopy (ARPES) measurements agree well with our first-principles results and thus establish Mg$_3$Bi$_2$ as an ideal materials platform for studying the exotic properties of type-II nodal line semimetals.

\end{abstract}

\maketitle

\section{Introduction}

The low-energy quasiparticles in semimetallic solid state systems provide an ideal platform for studying rich physics in elementary particles that has been theoretically proposed but never observed \cite{TI1, TI2}. The discovery of massless Weyl quasiparticles in Weyl semimetals represents a triumph of this approach \cite{WS1,WS2,WS3,WS4,WS5,WS6, WS7}. The exotic properties of Weyl semimetals such as chiral anomaly and non-local transports have stimulated a significant research interest in topological semimetallic phases of condensed matter. Recently, a new direction of thinking has emerged - to search for new topological quasiparticles that have no counterparts in high-energy physics. This idea offers the possibility for new topological phenomena that are not limited by the fundamental constraints that exist in high-energy physics. A prominent example showing this idea is the prediction and realization of type-II Weyl fermions \cite{WSII1,WSII2,WSII3,WSII4,WSII5,WSII6,WSII7}. The dispersion of Weyl quasiparticles in type-II Weyl semimetals is tilted along a given direction. This tilting of the low-energy quasiparticles explicitly breaks the Lorentz invariance, which is strictly required in high-energy physics, and gives rise to many distinctive properties such as anisotropic negative magnetoresistance \cite{weyl1, weyl2} and the existence of tilted surface bands \cite{weyl3}. Another novel topological solid-state phase is the nodal-line semimetal in which the intersection of conduction and valence bands consists of one-dimensional open or closed lines \cite{NL1,NL2, NL3, NL4, NL5}.  In contrast to the massless electrons of Weyl semimetals, the low-energy excitations of nodal-line semimetals are very massive along the direction tangent to the nodal lines and massless along the other two transverse directions in the momentum space. Apparently, there is no counterpart of such particles with extremely anisotropic mass in high-energy physics. This peculiar bulk band dispersion of nodal-line semimetals makes the density of states (DOS) of low-energy bulk excitations proportional to $|E-E_{F}|$, in contrast with the $(E-E_{F})^2$-like DOS in Weyl semimetals. Stronger electron correlation effects are expected in nodal-line semimetals because of the higher DOS at the Fermi energy. The nodal-line band structure is topological in the sense that the winding number along a little loop that interlinks the line nodes is nonzero. The nontrivial bulk band topology of nodal-line semimetals is accompanied by a two-dimensional surface band whose dispersion resembles a flat drumhead surface \cite{NL1, NL2, PbTaSe2}. The type-II Weyl and nodal-line semimetallic phases have been realized in many compounds such as WTe$_2$ \cite{WTe2, MoWTe2}, Ca$_3$P$_2$ \cite{Ca3P2}, Cu$_3$PdN \cite{Cu3PdN, Cu3PdN_2} and PbTaSe$_2$ \cite{PbTaSe2, Cava_PTS}. Combining the two ideas of type-II Weyl and nodal-line semimetals naturally leads to a new topological semimetallic phase named type-II nodal-line semimetal \cite{type2NL1, type2NL2}. In type-II nodal-line semimetals, the linear dispersion at every point of the nodal line is strongly tilted along one transverse direction in a similar way as in type-II Weyl semimetals, leading to remarkable differences in magnetic, optical, and transport properties compared with conventional nodal-line semimetals \cite{type2NL1}. Up to date, only one compound, K$_{4}$P$_{3}$, has been theoretically proposed as a type-II nodal-line semimetal but experimental evidence for its character is still lacking. Therefore, there is a pressing need for the realization of an experimental viable type-II Weyl semimetal to enable the study of the exotic behavior of this new type of topological matter.

In this work, we show by using first-principles calculations that Mg$_3$Bi$_2$ is a type-II nodal-line semimetal when spin-orbit coupling is ignored. The type-II nodal loop lies around $\Gamma$ slightly above the Fermi level, and the drumhead surface states appear inside the tilted nodal loop. With the inclusion of spin-orbit coupling, only a small energy gap ($\sim$ 35~meV) is opened at line nodes, yielding an overall bulk band structure that still resembles that of a type-II Weyl semimetal. Due to this prediction, we synthesized this compound in the form of large single crystals and carried out angle-resolved photoemission (ARPES) measurements. The ARPES results agree well with our first-principles band calculations, and thus establish Mg$_3$Bi$_2$ as a promising material for experimentally studying exotic type-II nodal-line semimetals.

\section{Results and discussions}

 %\section {Bulk band structure of Mg$_3$Bi$_2$}

Mg$_3$Bi$_2$ crystalizes in a La$_2$O$_3$-type structure in which Mg and Bi atoms form hexagonal layers and stack along $\mathbf{c}$ direction as shown in Fig.~1{\bf a}. The space group of the lattice is $P\bar{3}m1$ (No.~164) and the lattice is centrosymmetric. The crystal structure of our single-crystal Mg$_3$Bi$_2$ samples was determined by X-ray diffraction measurements, see Fig.~1{\bf b}, confirming their identity. The first Brillouin zone of the trigonal lattice is shown in Fig.~1{\bf c} with high-symmetry momentum points marked. The calculated bulk band structure of Mg$_3$Bi$_2$, obtained here using the generalized gradient approximation (GGA) method without the inclusion of spin-orbit coupling (SOC), is shown in Fig.~1{\bf d}. The low-energy electronic behavior of this compound is seen to be predominantly determined by the dispersion of two bands in the vicinity of the Fermi level. The bands cross each other along $\Gamma-K$ and $\Gamma-M$ directions and disperse apart in other directions, indicating a semimetallic property in this compound. The orbital projection (Fig.~1{\bf e}) indicates that the two bands originate from Bi-$6p$ and Mg-$3s$ orbitals. The band crossing in the $\Gamma-K$ direction is gapless and the two bands around the crossing point possess the same sign for their slope, which is the hallmark of a type-II Dirac point. Along $\Gamma-M$, by contrast, there exists a tiny gap between the two bands. Generally, in a spinless solid-state system with both space-inversion and time-reversal symmetries, the node points must form a one-dimensional line rather than discrete points, because the spinless Hamiltonian of two bands can be always taken to be real-valued and the dimension of nodal solutions is one in 3D momentum space \cite{Fang_spinless}. So in spinless Mg$_3$Bi$_2$, there must exist a type-II nodal line passing the K point. Our detailed band calculations demonstrate the nodal points indeed form a 1D loop surrounding $\Gamma$ and lie very close to the $\Gamma MK$ plane. The nodal loop does not stay entirely within the $\Gamma MK$ plane, but wiggles slightly with respect to the plane. The type-II feature holds on every point of the nodal loop. Therefore, Mg$_3$Bi$_2$ is a type-II nodal-line semimetal in the spinless case. In the presence of spin-orbit coupling, an energy gap is opened everywhere along the nodal line and the materials becomes a strong topological insulator because of this spin-orbit gap; see Figs.~1{\bf f} and 1{\bf g}. The typical size of this spin-orbit gap is 35~meV. Even though the type-II nodal line is gapped, the two bulk bands around the gap still share the same sign of slope, suggestive of a similar low-energy behavior of bulk carriers as the type-II nodal-line semimetals. 

%\section{Nodal loop and drumhead surface band of spinless Mg$_3$Bi$_2$}

The critical distinction between the type-I nodal line and the type-II nodal line is the band dispersion along transverse directions of the nodal line. In the type-I nodal-line band structure, the two bands disperse with opposite slopes in both transverse directions while in the type-II nodal-line band structure, the two bands are tilted in such a way that they share the same sign in their slope along one transverse direction. This is schematically plotted in Fig.~2{\bf a}. The zoom-in band structure of spinless Mg$_3$Bi$_2$ in Fig.~2{\bf b} clearly shows this characteristic of type-II nodal lines. The two bands cross at a nodal point (NP) in the $\Gamma-K$ direction ( $\parallelsum~k_x$) and form a tilted cone structure. Dispersing away from the NP point, the two bands have a quadratic touching along the tangent direction ( $\parallelsum~k_y$) of the nodal line and form a normal cone along the other transverse direction ( $\parallelsum~k_z$). We note that the band crossing is gaped along  the $\Gamma-M$ direction, so the nodal line is not entirely in the $\Gamma MK$ plane. To visualize the surface states, we calculated the band spectrum of a semi-infinite Mg$_3$Bi$_2$ slab with a (001) surface. The surface can be terminated by either a Bi atomic layer (Fig.~2{\bf c}) or an Mg atomic layer (Fig.~2{\bf d}). In both cases, there exist a surface band inside the nodal line as highlighted in the inset of Figs.~2{\bf c} and 2{\bf d}. This surface band connects to every bulk nodes along the nodal line, forming a 2D ``drumhead" surface. The ``drumhead" surface band is more dispersive on the Mg-terminated surface. 

When SOC is included, the bulk nodal ring is gapped, which results in a continuous gap between the conduction and valence bands throughout the Brillouin zone. This allows for a well-defined $\mathbb{Z}_{2}$ topological invariant \cite{Z2}. We calculate the $\mathbb{Z}_{2}$ topological invariant by the Wilson loop method \cite{Yu2011}; see Fig.~3{\bf a}. The Wilson band is an open curve traversing the entire Brillouin zone in the time reversal invariant plane $k_z = 0$ and a closed loop in another time reversal invariant plane $k_z = 0.5\pi$. The result indicates that the $\mathbb{Z}_{2}$ invariant equals 1, which indicates Mg$_3$Bi$_2$ is a strong topological insulator in the presence of SOC. The topological surface states are calculated by the semi-infinite Green's function. Each state is plotted with a color corresponding to the integrated charge density of the state within the top six atomic layers as depicted in Fig.~3{\bf b}. The thickness of the six atomic layers is about 10~$\mathrm{\AA}$, which roughly equals the typical escape depth of photoelectrons \cite{Depth}. Therefore, the weighted band spectrum of Mg$_3$Bi$_2$ (Figs.~3{\bf c-f}) can be directly compared to the ARPES results. For the Bi-terminated surface, a topological Dirac surface band connects the conduction and valence bulk bands (Fig.~3{\bf c}). The Dirac point is about 0.2 eV above the Fermi level, inside the spin-orbit gap. Below the Fermi level, there exist a pair of surface resonance bands around $\Gamma$, which are buried inside the bulk band region. The topological surface band on the Mg-terminated surface disperses deeper in energy with the Dirac point closer to the valence band edge (Fig.~3{\bf d}), unlike the Dirac point on the Bi-terminated surface, which is closer to the conduction band edge. The drumhead surface states in the spinless case have a similar energy position with respect to the bulk band edges. This is no surprise because the drumhead surface band splits and transforms into topological Dirac surface states when SOC is turned on. Another distinction between the band spectra on Bi- and Mg-terminated surfaces is that the resonance states on the Mg-terminated surface are more diffusive into the bulk and thus the calculated band spectrum is less sharp compared to those of Bi-terminated surface.  The calculated iso-energy band contours at $E=0.2$~eV are shown in Figs.~3{\bf e} and 3{\bf f} for Bi- and Mg- terminated surfaces, respectively. This energy corresponds to that of the Dirac point on the Bi-terminated surface. There are three prominent features in the band contour: a double lobe surrounding the $\bar{M}$ point, a little circular electron pocket centering at the $\bar{K}$ point and a little loop from the topological surface states at the zone center. The loop is vanishingly small on the Bi-terminated surface because the energy is set at the energy of the Dirac point.

To confirm the calculated band structure and thus the character of Mg$_3$Bi$_2$, we performed ARPES measurements on the single crystal samples. The spectrum taken along $\bar{K}-\bar{\Gamma}-\bar{K}$ with 74~eV photons is shown in Fig.~4{\bf a}. The dominant feature in the spectrum is a hole-like valence band pocket. Inside the bulk band region, there are three sets of surface states/resonances. The three sets of surface bands are all hole-like and the top of bands is at $E_F$, -0.4~eV and -0.8~eV, respectively. All these band features are well reproduced by the band simulation for the Bi-terminated semi-infinite slab (Fig.~4{\bf b}), indicating that the cleaved sample surface is Bi-terminated. We also carried out surface chemical doping to raise the Fermi level of the surface layers. Potassium deposition on the surface of Mg$_3$Bi$_2$ resulted in the Fermi level shift of 0.2~eV (Fig.~4{\bf c}); but further potassium deposition led to the deterioration of the spectra with no significant doping. The ARPES spectrum of the doped bands clearly shows the double humps at the top of the valence band above the Fermi level. The result is consistent with the calculated band structure. The projected Fermi surface contour as shown in Fig.~4{\bf d} consists of a hexagonal pocket centering at $\bar{\Gamma}$ and a lobe close to $\bar{M}$. We note that the intensity of the lobe at $\bar{M}$ is weak because the states are bulk like with a very small fraction of charge distributed within the top atomic layers. There is a small circular contour inside the $\bar{\Gamma}$ pocket, which is from the surface resonance bands. This is in excellent agreement with ARPES constant-energy maps of Fig.~4{\bf e}. The $\bar{M}$ pocket is missing from the ARPES data most likely because of the weak spectral weight within the surface layers. The size of both the bulk band pocket and surface state pocket increases as the energy changes from 0 to -0.6~eV, indicative of the hole-like nature of the bulk and surface bands.

To establish the nature of the measured states, we performed excitation energy-dependent ARPES mapping, Fig.~5{\bf a}. Using photons of 60--110 eV, which corresponds to $k_z$ values of 4.2 to 5.5 $\mathrm{\AA}^{-1}$, we show that the four inner sharp bands (marked by the four dashed lines in Fig.~5{\bf a}) have no $k_z$ dependence. This reflects the surface nature of those states and is consistent with the theoretical band structure shown in Fig.~3{\bf c}. The outer, broader bands do not appear to have much $k_z$ dependence either, but according to our calculations can only be ascribed to low-dispersing bulk states localized within the quadruple layers of Fig.~1{\bf a}. The two surface resonance bands form concentric circles in the iso-energy contour at the Fermi level as shown in Fig.~5{\bf b}. This double-circle surface Fermi surface is well reproduced by the first-principles simulation (Fig.~5{\bf b}). The topological character of Dirac surface states is further corroborated by inspecting the calculated spin texture of the surface band, which is shown in Figs.~5{\bf c-f}. The band spectra are plotted along $\bar{K}-\bar{\Gamma}-\bar{K}$ ($\parallelsum~k_x$) for the Bi surface termination (Fig.~5{\bf c}) and the Mg surface termination (Fig.~5{\bf d}), and the corresponding spin texture for the two surface conditions is reported in Figs.~5{\bf e} and 5{\bf f}, respectively. The calculated dominant spin component of the topological surface states is $\langle S_{y}\rangle$; the other two spin components are found to be negligible. Therefore, the calculated spin polarization of the Dirac surface states is perpendicular to the direction of the momentum, thus exhibiting a spin-momentum locking configuration, which is characteristic of topological surface states. The surface resonance bands also exhibit a similar helical spin texture with a dominant spin component $\langle S_{y}\rangle$. The two surface resonance bands on the Bi-terminated surface share the same spin helicity, indicating that, though the two bands share the same vertex point, they actually do not form a Rashba-type band structure.  The two branches of a Rashba spin-split band should have opposite spin helicities \cite{Rashba1, Rashba2}. 

\section{Conclusion}
 
We initially investigated the electronic band structure of Mg$_3$Bi$_2$ by first-principles calculations, motivated by the possibility that it might have topological electronic states. Single-crystal Mg$_3$Bi$_2$ samples were then grown and characterized by ARPES. The remarkable congruence of the ARPES measurements and the first-principles calculations unambiguously confirms the electronic band structure of Mg$_3$Bi$_2$. We find that Mg$_3$Bi$_2$ is a type-II nodal-line semimetal in the absence of SOC. The tilted nodal line is a closed loop centered at the $\Gamma$ point and the drumhead surface band is found inside the loop of bulk nodes. The spin-orbit interaction opens a band gap $\Delta E$ of 35~meV at the bulk line nodes and make the compound a strong topological insulator. The Dirac surface states in the topological insulator phase are derived from the drumhead surface states by incorporating SOC. We note that the energy gap of Mg$_3$Bi$_2$ is comparable to or smaller than the spin-orbit gap of typical type-I nodal-line compounds such as Cu$_3$PdN ($\Delta E =62$~meV) \cite{Cu3PdN, Cu3PdN_2}. The SOC present in Mg$_3$Bi$_2$ is mainly the result of the strong atomic SOC of Bi. Therefore, the spin-orbit band gap can be reduced by substitutional alloying, for example, by partially replacing Bi by Sb or As and making an Mg$_3$Bi$_{2-x}$Sb(As)$_x$ alloy. This method has been experimentally proven as effective to tune the band gap of topological materials \cite{Suyang2011}. Our theoretical and experimental results show that Mg$_3$Bi$_2$ is the archetype of a promising family of topological semimetals for studying the exotic properties of type-II nodal-line fermions and the associated drumhead surface states - a new topological electronic system that has thus far been inaccessible to experimental study.

\section{Methods}
%\subsection {Sample synthesis and XRD characterization}
Mg$_3$Bi$_2$ single crystals were grown out of a Bi-rich solution. Starting elements were loaded in alumina Canfield Crucible Sets\cite{CCS} with a molar ratio of Mg:Bi = 15:85, and sealed in a silica tube under partial argon atmosphere. The ampoule was then heated up to 600 $^{\circ}$C and slowly cooled to 300 $^{\circ}$C, at which temperature the crystals were separated from the solution in a centrifuge. Samples were analyzed for phase identification and purity using a Bruker D8 powder X-ray diffractometer equipped with Cu K$_\alpha$ radiation ($\lambda$=1.5406~$\mathrm{\AA}$). According to the XRD pattern shown in Fig.~1{\bf b}, Mg$_3$Bi$_2$  was successfully synthesized as the majority of the reflections were indexed according to the reported La$_2$O$_3$-type structure. 

%\subsection {ARPES measurements}
Photoemission data were collected at Advanced Light Source beamline 4.0.3 employing linearly polarized 60--110 eV photons, Scienta R8000 analyzer in dithered fixed mode, \textpm 15\textdegree{} angular lens mode, and polar angle scanning in steps of 1/3\textdegree{} for the perpendicular direction. The sample was attached to the cryostat, cleaved in ultrahigh vacuum at 35 K, and measured within 8 hours, showing linewidths of less than 0.025~$\mathrm{\AA}^{-1}$.

%\subsection {Computational methods}
We computed the electronic structures using the projector augmented wave method \cite{DFT1,DFT2} as implemented in the VASP package \cite{DFT3} within the generalized gradient approximation schemes \cite{DFT4}. The SOC was included self-consistently in the calculations of electronic structures with a Monkhorst-Pack $k$-point mesh 15 $\times$ 15 $\times$ 11. To calculate the surface electronic structures, we constructed first-principles tight-binding model Hamilton by projecting onto the Wannier orbitals \cite{DFT5, DFT6, DFT7}, which use the VASP2WANNIER90 interface \cite{DFT8}. We use Mg $s$ and $p$ orbitals and Bi $p$ orbitals to construct Wannier functions and without perform the procedure for maximizing localization. The surface states are calculated from the surface Green's function of the semi-infinite system.

Note: during the preparation of this manuscript, we noticed that similar band structure calculation of Mg$_3$Bi$_2$ has been presented in a theoretical paper \cite{Zhang17}. It was reported theoretically in Ref.\cite{Zhang17} that Mg$_3$Bi$_2$ is a gapless Dirac semimetal in the presence of SOC. This result is in error. According to our calculations, Mg$_3$Bi$_2$ is a topological insulator with a tiny band gap. Our theoretical result is in good agreement with the ARPES measurements on Mg$_3$Bi$_2$ single-crystal samples.

\section{Acknowledgements}

\newpage

\begin{figure*}
\centering
\includegraphics[width=16cm]{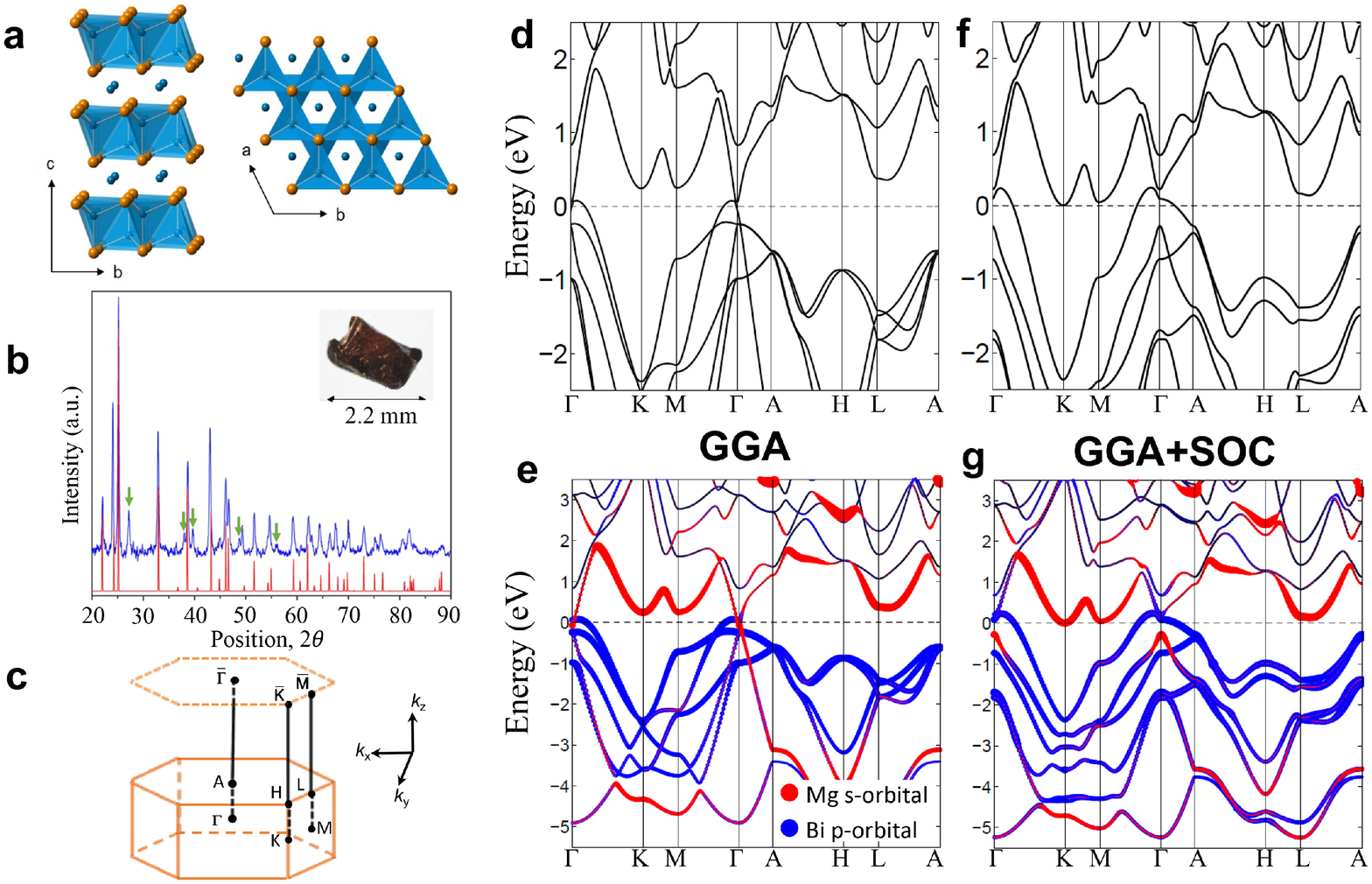}
\caption{{\bf Crystal structure and bulk bands of Mg$_3$Bi$_2$.} {\bf a}, The lattice structure of Mg$_3$Bi$_2$ (Blue: Mg; Orange: Bi). {\bf b}, Powder X-ray diffraction (PXRD) pattern. The green arrows correspond to the reflections of remaining Bi flux on the sample surface. {\bf c}, Bulk Brillouin zone and projected (001)-surface Brillouin zone. {\bf d}, Bulk band structure without the inclusion of spin-orbit coupling. {\bf e}, Bulk bands in {\bf d} with atomic orbital projection. {\bf f} and {\bf g}, Same as {\bf d} and {\bf e}, but with the inclusion of spin-orbit coupling.}
\label{fig1}
\end{figure*}

\newpage

\begin{figure*}
\centering
\includegraphics[width=13cm]{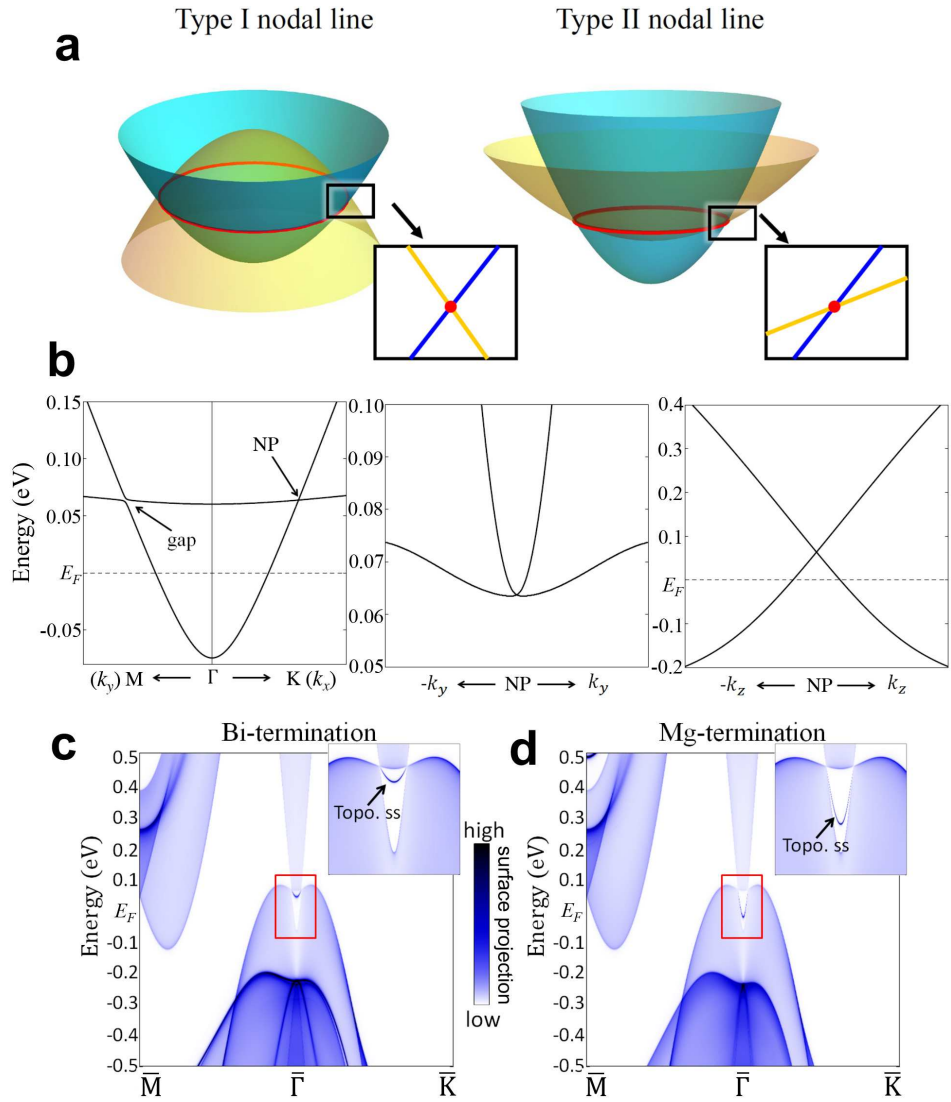}
\caption{{\bf Type-II nodal line band structure of Mg$_3$Bi$_2$ (without SOC).} {\bf a},  Schematic of type-I and type-II nodal line band structure. {\bf b}, Bulk bands around a nodal point (NP) in $\Gamma K$ direction. Three panels show the band dispersion from the NP in three perpendicular (radial, tangential, and vertical) directions, respectively. {\bf c}, Band spectrum of a semi-infinite slab with a (001) surface. The surface is terminated at a Bi layer. The surface band is shown in the inset. The color indicates the weight of charge density of each state at the surface layer. {\bf d}, Same as {\bf c}, but for a Mg-terminated semi-infinite slab.}
\label{fig2}
\end{figure*}
%
%\newpage
%
\begin{figure*}
\centering
\includegraphics[width=13cm]{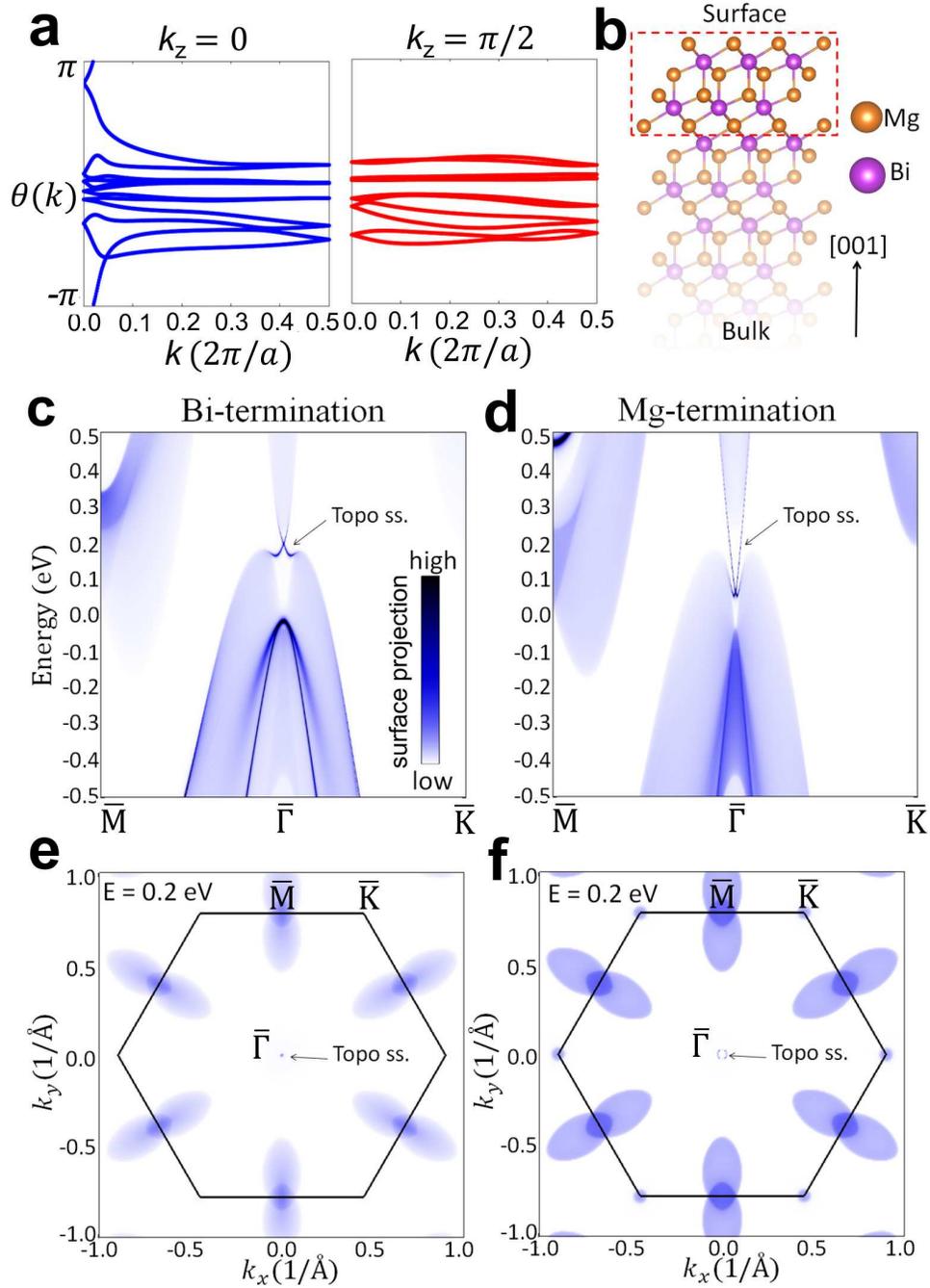}
\caption{{\bf Topological invariant and surface band structure of Mg$_3$Bi$_2$ (with SOC).} {\bf a}, Wannier charge center evolution in time-reversal invariant planes. {\bf b}, Lattice structure of a semi-infinite slab. Band spectrum of a semi-infinite slab with a Bi-terminated (001) surface. The color indicates the weight of charge density of each state within the top six atomic layers. {\bf d}, Same as {\bf c}, but for a Mg-terminated semi-infinite slab. {\bf e} and {\bf f}, Calculated Iso-energy band contour at $E=0.2$~eV for Bi-terminated and Mg-terminated semi-infinite slabs, respectively.}
\label{fig3}
\end{figure*}
%
%
%\newpage
%
\begin{figure*}
\centering
\includegraphics[width=16cm]{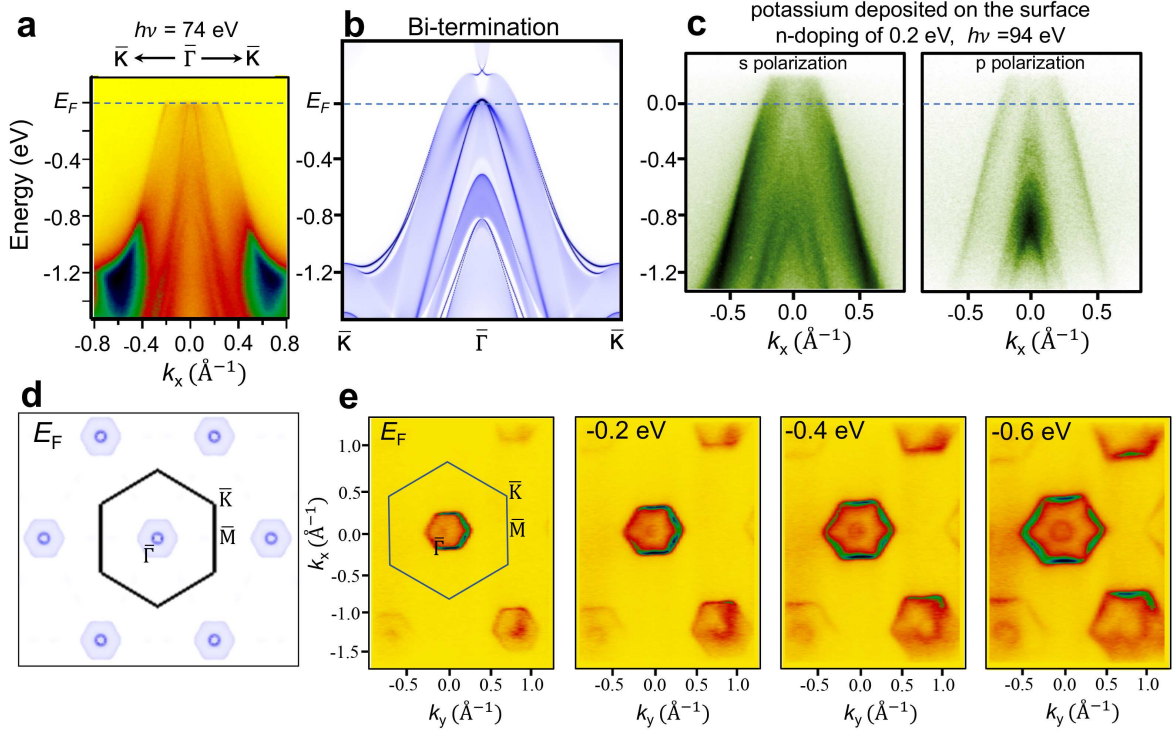}
\caption{{\bf ARPES measurements on Mg$_3$Bi$_2$.} {\bf a}, ARPES spectrum taken along $\bar{\Gamma}-\bar{K}$ direction. the photon energy is 74~eV. {\bf b}, Calculated band spectrum of a semi-infinite slab with a Bi-terminated (001) surface. {\bf c}, APRES spectra taken after potassium deposition of the surface. The two panels show the ARPES results taken with vertically and horizontally polarized photons, respectively. {\bf d}, Calculated Iso-energy band contour at the Fermi level for a Bi-terminated (001) surface. {\bf e}, Iso-energy ARPES mapping at different binding energies.} 
\label{fig4}
\end{figure*}
%
%\newpage
%
\begin{figure*}
\centering
\includegraphics[width=16cm]{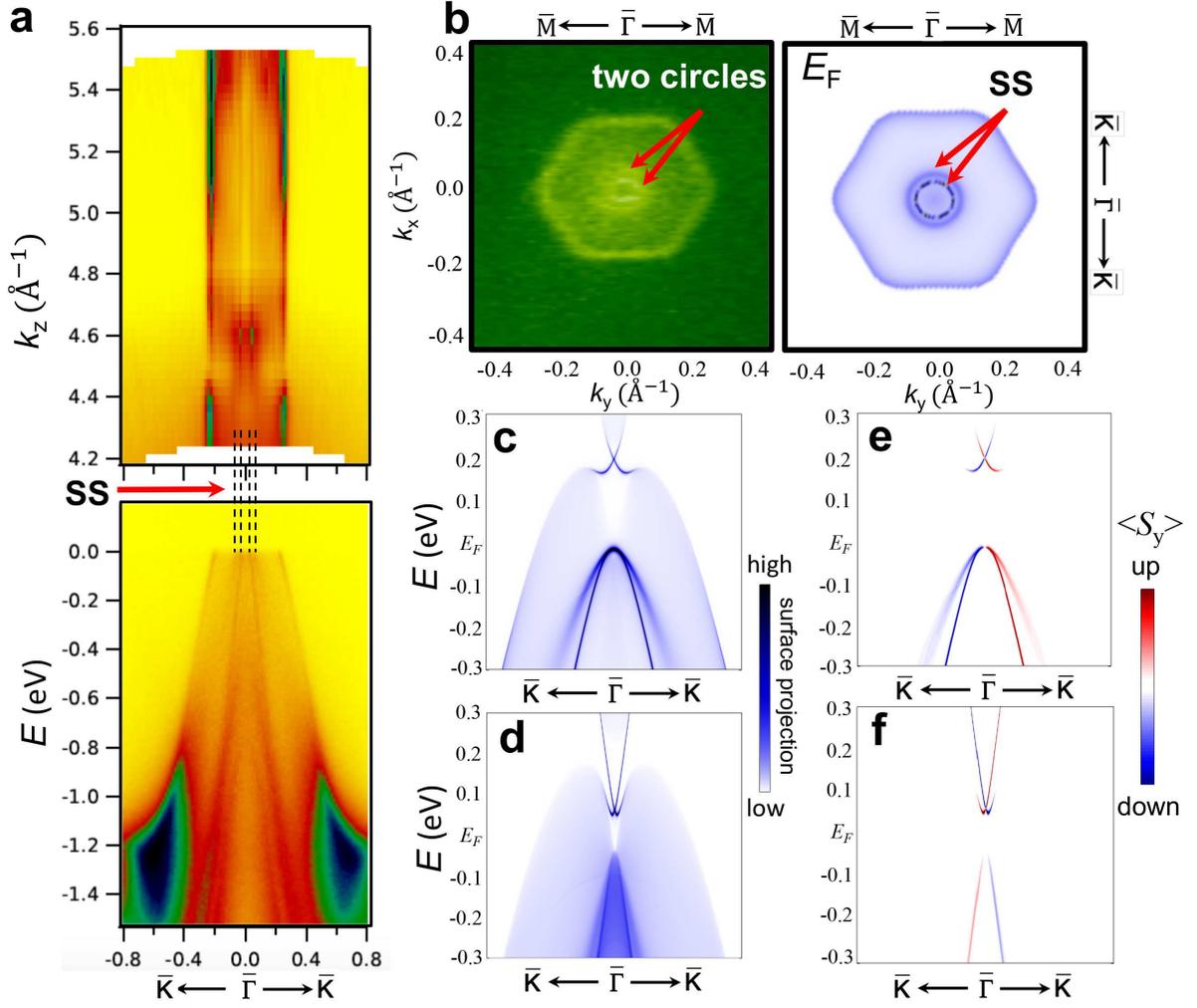}
\caption{{\bf ARPES results of surface states of Mg$_3$Bi$_2$.} {\bf a}, Momentum distribution curves at the Fermi level taken with different photon energies (and thus different $k_z$ values). {\bf b}, ARPES and calculated iso-energy contour at the Fermi level. The two inner circles are the surface states. {\bf c} and {\bf d}, Calculated band spectra along $\bar{\Gamma}-\bar{K}$ direction for Bi-terminated and Mg-terminated semi-infinite slabs, respectively. The color indicates the surface weight of each state. {\bf e} and {\bf f}, Calculated spin texture for Bi-terminated and Mg-terminated semi-infinite slabs, respectively. The color indicates the spin-polarization along $S_y$ direction.}
\label{fig5}
\end{figure*}

\end{document}